\documentclass[10pt,conference,letterpaper]{IEEEtran}

\usepackage{mathptmx}
\usepackage[ruled,vlined]{algorithm2e}
\usepackage{booktabs}
\usepackage{color}
\usepackage{graphicx}
\usepackage{listings}
\usepackage{multirow}
\usepackage[all]{nowidow}
\usepackage{subcaption}
\usepackage{url}

\usepackage{hyperref}
\usepackage{breakurl}
\usepackage{flushend}
\usepackage{paralist}
\let\labelindent\relax
\usepackage{wrapfig}
\usepackage{lipsum}
\usepackage[inline]{enumitem}
\usepackage[font=footnotesize,belowskip=-8pt]{caption}
\usepackage[numbers,sort&compress]{natbib}

\lstdefinestyle{json_cert}{
basicstyle=\footnotesize\fontfamily{lmtt}\fontseries{l}\selectfont,
string=[s]{"}{"},
alsoletter= 0123456789,
morecomment=[s]{subject}{:},
morecomment=[s]{issuers}{:},
morecomment=[s]{notBefore}{:},
morecomment=[s]{notAfter}{:},
morecomment=[s]{publicKey}{:},
morecomment=[s]{schnorrSignature}{:},
morecomment=[s]{transaction}{:},
morecomment=[s]{blockNo}{:},
morecomment=[s]{inclusionProof}{:},
commentstyle=\bfseries,
captionpos=b,
columns=fullflexible,
showstringspaces=false
}

\lstdefinestyle{contract_pseudo}{
basicstyle=\footnotesize\fontfamily{lmtt}\fontseries{l}\selectfont,
keywords={contract, function, if, else, new, input, output, struct, for, in},
keywordstyle=\bfseries,
captionpos=b
}

\newcommand{\name}{BlockPKI\xspace}
\newcommand{\myparagraph}[1]{\noindent{\bf #1.}}

\begin{document}

\title{\name: An Automated, Resilient, and Transparent Public-Key
Infrastructure}

\author{\IEEEauthorblockN{Lukasz Dykcik}
\IEEEauthorblockA{\textit{ETH Zurich}}
\and
\IEEEauthorblockN{Laurent Chuat}
\IEEEauthorblockA{\textit{ETH Zurich}}
\and
\IEEEauthorblockN{Pawel Szalachowski}
\IEEEauthorblockA{\textit{SUTD}}
\and
\IEEEauthorblockN{Adrian Perrig}
\IEEEauthorblockA{\textit{ETH Zurich}}
}

\maketitle

\begin{abstract}
This paper describes \name, a blockchain-based public-key infrastructure that
enables an automated, resilient, and transparent issuance of digital
certificates. Our goal is to address several shortcomings of the current TLS
infrastructure and its proposed extensions. In particular, we aim at reducing
the power of individual certification authorities and make their actions
publicly visible and accountable, without introducing yet another trusted third
party. To demonstrate the benefits and practicality of our system, we present
evaluation results and describe our prototype implementation.
\end{abstract}

\section{Introduction}
\label{sec:introduction}

Now more than ever, the use of secure communication protocols is encouraged and
promoted. Browser vendors (e.g., Google with Chrome~\cite{ChromeIndicator2016},
or Mozilla with Firefox~\cite{FirefoxIndicator2016}) have started changing the
appearance of security indicators in the address bar: regular HTTP connections,
which used to look neutral~\cite{felt2016rethinking}, are now described as ``not
secure'', whereas HTTPS connections are clearly labeled as ``secure''. As more
and more administrators and developers are becoming aware of the risks
associated with using unsecure protocols, and as vulnerable means of
communication are being deprecated~\cite{Fisher2015Apple}, cryptographically
protected protocols (such as TLS) are becoming the norm.

Unfortunately, the TLS public-key infrastructure (PKI) suffers from a
weakest-link security problem: any trusted certification authority (CA) can on
its own produce a valid certificate for any domain name. A certification
authority is considered trusted by a client if its certificate is present in the
client's list of root CAs or signed by another trusted CA. By compromising any
root or intermediate CA, an attacker can compromise the security of the entire
system. For this reason, Google developed the Certificate Transparency (CT)
framework~\cite{rfc6962}, which relies on append-only logs to make certificates
publicly available. Unfortunately, that approach comes with a few drawbacks: CT
enables the detection of CA misbehavior but does not prevent it. Moreover, the
list of log servers is maintained by a single entity. Although the log's
contents can be consulted and proved to be consistent, log servers can choose to
ignore queries. Finally, to avoid a split-world attack, in which a malicious log
server would show inconsistent versions of the log to different clients, a
gossip protocol is needed~\cite{chuat2015efficient}. Consequently, in order to
tolerate a malicious or compromised CA, every certificate issuance should
involve multiple CAs and all operations should be logged securely and in a fully
distributed way.

For TLS to become truly ubiquitous on the web, the issuance of digital
certificates must be frictionless. In this spirit, the ``Let's Encrypt'' CA
offers free, widely-accepted certificates as well as tools for automating their
issuance, re-issuance, and revocation. These unique characteristics have made it
incredibly successful, with over 50 million active certificates as of March
2018~\cite{LetsEncryptStats}. Let's Encrypt has managed to offer free services
only thanks to a multitude of sponsors, though; automation helps reducing costs,
but it does not eliminate them~\cite{letsencrypt_cost}. Therefore, an ideal PKI
(which involves multiple CAs for each certificate issuance) should also include
a payment framework, so that non-subsidized CAs can be remunerated efficiently
and conveniently.

In this paper, we present \name, a public-key infrastructure that employs
smart contracts to provide the following features:
\begin{compactitem}
	\item \textbf{Automation:} The certificate issuance process---including the
	transmission of a request to selected CAs, proving control over the
	domain name, making the newly created certificate publicly visible, and
	paying the signing CAs---necessitates minimal human involvement.
	\item \textbf{Resilience:} No single entity has the ability to
	issue an illegitimate certificate or disrupt normal operations.
	\name relies on a blockchain, which prevents the situation where
	a single entity (such as a log server) can ignore requests and block the
	issuance or verification process. Moreover, several CAs must be involved
	to produce a valid certificate, which greatly
	improves the system's tolerance to compromise.
	\item \textbf{Transparency:} All operations relative to certificate
	management are logged in the blockchain, which allows monitoring of
	all operations and detection of anomalies.
\end{compactitem}

To achieve similar objectives, many existing schemes introduce various trusted
parties (such as log servers) with different roles, scopes, and
responsibilities~\cite{kim2013aki,basin2014aar,syta2016keeping,merzdovnik2016whom}.
A blockchain, as a distributed append-only ledger, can support a certificate log
without introducing any trusted third party. Using a PKI-dedicated gossip
protocol becomes unnecessary; exchanging and storing block headers is sufficient
for clients to verify that all relevant operations are logged consistently.

\section{Background}
\label{sec:background}

\name makes use of several existing data structures and schemes that we briefly
describe in this section.

\subsection{Schnorr Multi-Signatures}
\label{schnorr}

The Schnorr algorithm~\cite{schnorr1989efficient} is a digital signature scheme
that can be extended~\cite{okamoto1999multi} to a multi-signature scheme,
enabling several independent entities to produce a concise signature. Creating a
multi-signature involves two rounds. Consider $n$ entities signing a statement
$m$.  Each entity is characterized by its public key $Q_i$ and private key
$x_i$, where $Q_i=g^{x_i}$ and $g$ is a publicly known group generator. In the
first round, each entity generates a pair of private/public nonces ($k_i$,
$N_i$), such that $N_i=g^{k_i}$, and communicates the public nonce $N_i$ to all
other participants. In the second round, each entity combines all public nonces
and gets $\overline{N} = \prod_i^nN_i$.  With $\overline{N}$ and its private
nonce $k_i$, the entity computes its partial signature $s_i = k_i-ex_i$, where
$e=h(\overline{N} \parallel m)$ and $h$ is a hash function, and broadcasts it.
Now anyone can combine all partial signatures $\overline{s} = \sum_i^ns_i$ to
form a merged signature ($e$, $\overline{s}$), which can be verified with a
combined public key $\overline{Q} = \prod_i^nQ_i$, similarly to a single Schnorr
signature, i.e., by calculating $\overline{N}' = g^{\overline{s}}\overline{Q}^e$
and checking whether $e=h(\overline{N}'\parallel m)$. Combining public keys or
partial signatures does not require knowledge of any signer's private key. The
private nonce must be kept private and destroyed after creating the partial
signature as disclosing it may leak the long-term private key $x_i$.

\subsection{Blockchains and Smart Contracts}
\label{background:blockchain}

A smart contract is a computer program executed in a trusted, impartial
environment. In contrast to real-life agreements, no third parties are needed to
enforce the rules written in the contract. Currently, the most common way of
ensuring an impartial execution environment for smart contracts is to use a
blockchain. A blockchain is a data structure where each block header contains a
cryptographic hash of the previous block. Each block also contains a list of
transactions.

Blockchains typically rely on \emph{Merkle trees} to efficiently and securely
include transactions in block headers. In a Merkle tree, each non-leaf node is
labeled with the hash of all its child nodes. Such a data structure allows
anyone to produce concise, unique, and easy-to-verify \textit{inclusion proofs}.
The size of such a proof grows logarithmically with the number of leaves in the
tree. The advantage of this approach is that even a so-called \emph{light
client} (i.e., clients holding only block headers, rather than the entire
blockchain) can verify that a given transaction is included in the blockchain.

Typically, new blocks (with new transactions) are appended to the blockchain in
a stochastic process called \emph{mining}; every block must be backed up by
sufficient \textit{proof of work}, i.e., finding a new block requires solving a
computationally intensive problem. It may happen that two miners produce a proof
of work for the next block simultaneously, then two versions of the blockchain
may coexist. This problem is resolved by the rule that the longest chain is
accepted as the valid one. Having these properties, one can treat the blockchain
as a distributed database in which every block is immutable and creating an
alternative version of the entire database is computationally infeasible due to
the above-mentioned proof-of-work mechanism.

With smart contracts running on a blockchain, censorship and manipulations
become highly impractical. Every machine that participates in the blockchain
system by checking the validity of new blocks is called a \textit{node}, some
nodes are also miners. Machines that download and verify block headers only and
acquire  information about transactions on an as-needed basis are called
\textit{light clients}. Blockchain users have accounts identified by unique
addresses. Accounts can be used to publish new contracts or interact with
existing ones. Each contract running on the blockchain is also identified by a
unique address. To call a method in a contract, the user needs to know the
method's name and arguments.

Usually, blockchain systems are open and prevent spamming by charging a small
fee for each action. In order to be included in the blockchain each transaction
must be accompanied by a fee. The fee is then transferred to the miner of the
block with this transaction. Fees are expressed using a cryptocurrency, a
digital token with a monetary value. Cryptocurrency can be mined, received from
another user, or bought with fiat money on specialized exchange platforms.

\subsection{Automatic Certificate Management}
\label{acme}

The Automatic Certificate Management Environment or
ACME~\cite{barnes2015automatic} is a proposed standard used by Let's Encrypt to
automate the issuance of domain-validated (DV) certificates. The domain
validation process of ACME involves three major steps:
\begin{enumerate*}[label=(\alph*)]
    \item Upon request, the CA sends a set of challenges  to the purported
        domain owner. These challenges can consist in setting up a special DNS
        record or making a file with a specified content available at a given
        path under the requested domain name. 
    \item A script completes the challenges and sends to the CA a set of
        responses specifying which challenges have been completed. 
    \item The CA checks whether the challenges have been correctly executed and
        issues the certificate. 
\end{enumerate*}
Note that being able to control the domain name or the web server's root
directory suffices to obtain a DV certificate.

\section{System Model}

We distinguish four main roles in \name:

\begin{compactitem}
    \item \textbf{Requesters} control domain names and want to obtain
        certificates by proving ownership of these domain names to CAs. 
    \item \textbf{Certification authorities (CAs)} in \name are similar to
    	today's root certification authorities. They are trusted entities (i.e.,
    	their public keys are axiomatically trusted by the clients' software)
    	able to accept requests, verify the identity of the requester, and
    	produce a signature.
    \item \textbf{Clients} want to establish secure TLS connections to web
        servers. They check the validity of the delivered certificate
        using a list of self-signed root CA certificates, which usually comes
        with the client's browser or OS. There are three types of clients, with
        varying security guarantees (from the weakest to the strongest):
        
        \begin{compactitem}
        \item \textbf{Blockchain-unaware:} This type of client does not interact
        with the blockchain whatsoever, and thus cannot verify that a received
        certificate is visible to everyone (so that a potential anomaly can be
        detected), but still benefits from the security advantage of
        multi-signatures.
        \item \textbf{Lightweight:} Such a client fetches, validates, and stores
        block headers to verify that a received certificate corresponds to a
        transaction that is included in a block.
        \item \textbf{Full node:} Clients that maintain the complete blockchain
        are called \emph{full nodes}. Doing so requires substantially more
        resources than only fetching block headers but provides the maximum
        level of security.
        \end{compactitem}
        
	\item \textbf{Web servers} are controlled by requesters and can be used to
	    prove ownership of a domain name, i.e., if a domain name resolves to the
	    address of a web server that the requester controls, then the requester
	    can make it serve any content from any given path as evidence of
	    ownership. The goal of the requester is to obtain a certificate to
	    secure connections from clients to its web server.
\end{compactitem}

Throughout the paper, we assume the following:
\begin{enumerate*}[label=(\alph*)]
\item The public keys and blockchain account addresses of root CAs are publicly
        known.
\item The code of \name contracts is publicly known, thus everyone can verify
and audit the code and everyone knows how to interact with these contracts.
    \item The blockchain addresses of permanent contract instances are known to
        all requesters and CAs.
\end{enumerate*}

For the sake of simplicity, we also assume that CAs and requesters are locally
maintaining the blockchain. However, requesters could use a proxy to interact
with the blockchain.

A parameter $T$ specifies the number of CAs signing the
certificate.  The adversary's goal is to create a fraudulent
certificate for a domain.  To this end, the adversary can compromise CAs and
conduct impersonation attacks, but cannot compromise the targeted domain.
Specifically, we assume that the adversary can corrupt $i$ CAs and mount a
domain-impersonation attack on $j$ other CAs, as long as $i+j<T$.
If $i+j\geq T$ the adversary can create a fraudulent certificate that will be
accepted by clients. Even in such a case, however, \name's objective is to
provide a high probability of attack detection.

Furthermore, we assume that the attacker does not control the majority of the
blockchain computational power. This means that she cannot launch a so-called
\textit{51\% attack} on the underlying blockchain, thus she cannot roll back
changes in the blockchain and cannot censor transactions from entering the
blockchain.

\section{\name Overview}

In \name, CAs conduct automated domain validation. Moreover, in order to
address the weakest-link problem of
the current infrastructure, \name mandates multi-path probing (i.e.,
each domain name must be validated by several CAs).
One of our goals is to automate not only the domain-validation process, but
also the certificate creation process. To this end, \name relies on smart
contracts,
which allow automated interactions between requesters and CAs.

\name requires that each certificate be signed by $T$ CAs,
so that one careless or compromised CA cannot create client-accepted
certificates by itself. Requesters can choose the value of $T$ they want
to use, but browser vendors will effectively dictate a lower bound on this
value, as browsers may refuse a certificate signed by an insufficient number
of CAs. We envision that $T$ will initially be very small (perhaps even
just one during early stages of deployment) and then increase over time.

In this section, we provide a brief overview of how \name deals with two main
aspects
of public-key infrastructures: certificate issuance and certificate
verification.

\begin{figure}[t]
    \centering
    \includegraphics[width=\columnwidth]{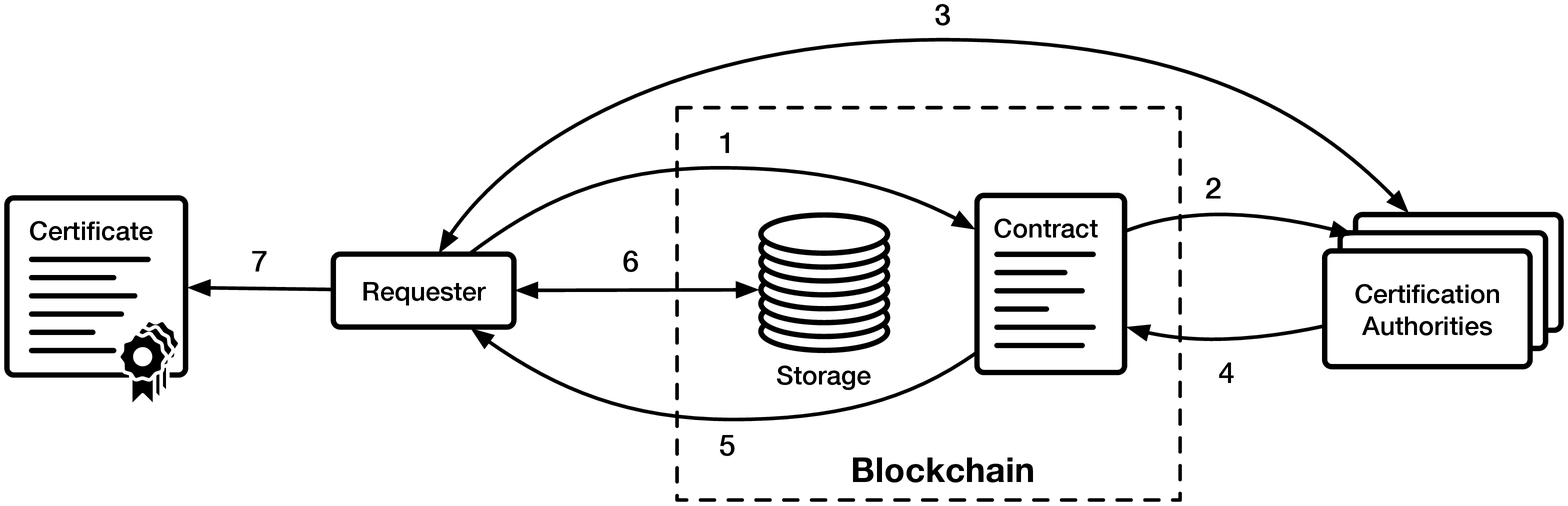}
    \caption{Overview of the certificate issuance process.}
    \label{overview}
\end{figure}

The certificate issuance process is illustrated in Figure~\ref{overview}
and summarized below:
\begin{compactenum}
    \item The requester creates a contract on the blockchain that will gather
signatures from CAs. In the contract, the requester specifies certificate
        parameters, the blockchain addresses of $T$ CAs authorized to sign the
        certificate, and the amount of financial compensation for each
        signature.
    \item The CAs detect the new contract and check whether they are authorized
        to validate the public key of the requester.
    \item The authorized CAs perform the public-key verification. To pass the
        verification, the requester must prove control of the domain
        to the CAs. 
\item If the verification is successful, each authorized CA sends its partial
        signature to the contract and receives the financial compensation.
    \item The requester gathers signatures from the blockchain, combines them
        into a single multi-signature and appends it to the certificate data.
    \item The requester puts the multi-signature and certificate data as a
        payload in a transaction and publishes it in the blockchain.
    \item The transaction together with the proof of its inclusion in a mined
        block forms a \name certificate.
\end{compactenum}

Certificates are used for establishing TLS connections. When a client connects
to a server, the clients obtains the domain's certificate during the TLS
handshake.
In \name, to validate the certificate's authenticity the client verifies
\begin{enumerate*}[label=(\alph*)]
    \item that the domain name is correct and that the certificate is
        not expired,
    \item whether the certificate is signed by
        the threshold number $T$ of CAs (trusted by the client), and
    \item whether the locally maintained block headers or blockchain match
    	the inclusion proof of the transaction containing the certificate
    	data (if the client is either a lightweight or full blockchain node).
\end{enumerate*}

\section{\name in Detail}
\label{sec:details}

\subsection{Contracts}

Smart contracts are used for the automation of most interactions:
certificate request, payment handling, and certificate
issuance. \name uses three types of contracts:
\begin{compactitem}
    \item \textbf{The central contract} receives certificate requests, verifies
        them, and maintains a list of accepted requests. There is a
        single instance of the central contract on the blockchain, and all
        requesters and CAs know its blockchain address.
    \item \textbf{Domain contracts} are created by the central contract and
        serve certificate requests (i.e., they receive signatures and issue
        compensation). A new domain contract is created for every certificate
        request.
    \item \textbf{The storage contract} is used as a recipient of transactions
        containing certificate data and a Schnorr signature.
\end{compactitem}

\myparagraph{Central Contract}
To obtain a certificate, the requester must create a new domain
contract using the central contract, which works as a contract factory. The
pseudocode of the central contract is presented in
Listing~\ref{centralContract} in the appendix.
The central contract ensures that all certificate requests are stored in one
place in the blockchain and that each domain contract has the same, publicly
known code, so that protocol parties have a uniform way to interact with each
other. Furthermore, the central contract guarantees that all domain
contracts are accompanied with sufficient funds to pay all CAs, and that
provided
parameters are valid.
The central contract implements a method called
\texttt{create\-Domain\-Contract}
that requesters use to create a domain contract with their parameters.
The requester specifies the following parameters: certificate data, the list of
authorized CAs (identified by blockchain addresses), and the amount of financial
compensation for each signature.
If the amount of cryptocurrency supplied when creating a new
domain contract is insufficient to cover all compensations, then the contract
cannot be created.  For flexibility, a
domain contract can handle any value of $T$. However, we assume that $T$ is a 
global system parameter (i.e., browser and requester software should all use the
same value), but it may be increased in the long-run for improved security.
A requester may also list more CAs than $T$ for increased resilience, as we
discuss in Section~\ref{sec:discussion}, but we ignore this possibility
for the time being for the sake of simplicity.

\myparagraph{Domain Contract}
The domain contract is created for every certificate request
and is a crucial element in automating certificate issuance. 
It enables CAs to create a multi-signature by providing a platform
for exchanging public nonces.
In particular, the domain contract handles the following operations required to
issue a certificate:
\begin{enumerate*}[label=(\alph*)]
    \item receiving $T$ CA signatures over the certificate, and
    \item paying the CAs for submitting their signatures.
\end{enumerate*}
Additionally, the contract implements the logic ensuring that the protocol is
executed in the correct order (e.g., it receives signatures after all nonces are
present).  The code of the domain contract is presented in
Listing~\ref{domainContract}. A new domain contract is constructed with the
parameters provided when calling the \texttt{create\-Domain\-Contract} method.
The contract contains all the information necessary for obtaining a certificate:
the domain name, the public key, and the validity period. Additionally, the
requester specifies the CAs allowed to participate in the issuance of the
certificate and sets a compensation for each of the CAs.

\myparagraph{Storage Contract}
The contract serves as a recipient of transactions containing signed certificate
data. Upon reception of the transaction, the contract stores its content in an
internal storage. Pseudocode of the  contract is provided in
Listing~\ref{storageContract}.

\subsection{Certificate Issuance}

CAs observe the blockchain for new domain contracts.  Once a new contract
is detected, each CA checks if it is specified as authorized to validate
the domain's public key.

\myparagraph{Domain Validation}
If the CA is requested to issue the certificate, it performs an ACME domain
validation. In this process, the owner of the public key proves ownership of the
domain, hence the association between the public key and the domain can be
established.  The requester's server runs a program that waits for and completes
all verification challenges sent from CAs. The challenges, for example, may
require to add a file with a specified content at a specified path on the web
server. If verification confirms the validity of the information in the domain
contract, the CA proceeds to signing the certificate.

\myparagraph{Signing the Certificate}
To keep certificates compact, we use the Schnorr algorithm in the
multi-signature setting (see Section~\ref{schnorr}). The signing process begins
with generating a private-public nonce pair ($k_i,N_i$) and sending the public
nonce $N_i$ in a transaction to the domain contract (see
\texttt{sendCertPubNonce} in Listing~\ref{domainContract}). After all $T$ nonces
are put in the domain contract, an authorized CA can send its signature $s_i$ by
calling \texttt{sendCertSignature}. To issue the signature, nonces of all other
CAs involved in creating the Schnorr multi-signature are needed. The CA extracts
them from the domain contract and combines them. Then, the CA sends its partial
signature of the certificate to receive its compensation. All partial signatures
are permanently stored in the domain contract, they can be read by any
blockchain user and combined together into a single, compact Schnorr signature.

\myparagraph{Creating the Final Certificate}
Gathering signatures is one of the main goals of the domain contract.  The
requester observes her domain contract and waits until all signatures are
gathered. Then, the requester combines all partial signatures into a compact
Schnorr multi-signature. If the signature is correct, the requester creates a
list of IDs of CAs that took part in signing the certificate; the IDs will
enable clients to verify the signature. Then, the requester sends a transaction
with the certificate data, the list of IDs, and the multi-signature to the
storage contract and waits for the inclusion of his transaction in a block in
the blockchain. Once a block with the transaction is mined, the requester reads
the inclusion proof of the transaction in the Merkle tree encompassing all
transactions included in the block. The root of this tree is contained in the
block header. The transaction together with its inclusion in a particular block
creates a certificate as presented in Listing~\ref{certificate}. To make sure
that the transaction is included in the blockchain and that the chain will not
be re-written, the requester should wait until a number of blocks are mined.
With each block mined on top of the one with the considered transaction, chances
of the transaction being removed from the blockchain exponentially decrease
to~0~\cite{vitalikForks}. After the certificate is created, the requester can
upload it to the domain's servers, so that it can be used to establish TLS
connections with clients.

\begin{figure}
\begin{lstlisting}[mathescape, style=json_cert, frame=single,
caption=Example of \name certificate., label=certificate]
$\textrm{\small Certificate data (similar to X.509), logged in the blockchain:}$

 { subjectName: "www.example.com",
   issuers: ["CA1", "CA2", "CA3", "CA4"],
   notBefore: 1 January 2018, 01:00:00 CEST,
   notAfter: 1 February 2018, 01:00:00 CEST,
   publicKey: B1:E1:37...,
   schnorrSignature: 4A:BD:FF...,
   ... }

$\textrm{\small Domain certificate with inclusion proof:}$

 { transaction: C5:93:83..., // contains certificate data
   blockNo: 123456,
   inclusionProof: [AB:99:7F..., 63:D5:F7...
     AB:D6:9A..., F4:23:89..., A7:45:63..., ...] }
\end{lstlisting}
\end{figure}

\subsection{Certificate Verification} \label{section_certificate_verification}

The certificate verification process is presented in
Listing~\ref{cert_verification} (see Appendix).
Certificate data, a Schnorr multi-signature, and the list of CAs that
contributed
to the multi-signature are contained in the payload of the transaction delivered
as a part of the certificate (see Listing~\ref{certificate}). First, the client
reads data from the transaction and combines public keys of the CAs that are
listed in the transaction (i.e., in the certificate). The clients also make
sure that signing CAs are trusted.
Then, the client ensures that
\begin{enumerate*}[label=(\alph*)]
    \item the domain name from the certificate matches the visited website's
        address,
    \item the certificate is not expired, 
    \item the multi-signature over the certificate is correct (i.e., signed by
        the listed CAs), and
    \item the transaction corresponding to the certificate is present in the
        blockchain.
\end{enumerate*}
We note, however, that this last operation can be performed asynchronously and
only by clients who maintain either a lightweight of full version of the
blockchain.

\begin{figure}
    \centering
    \includegraphics[width=\columnwidth]{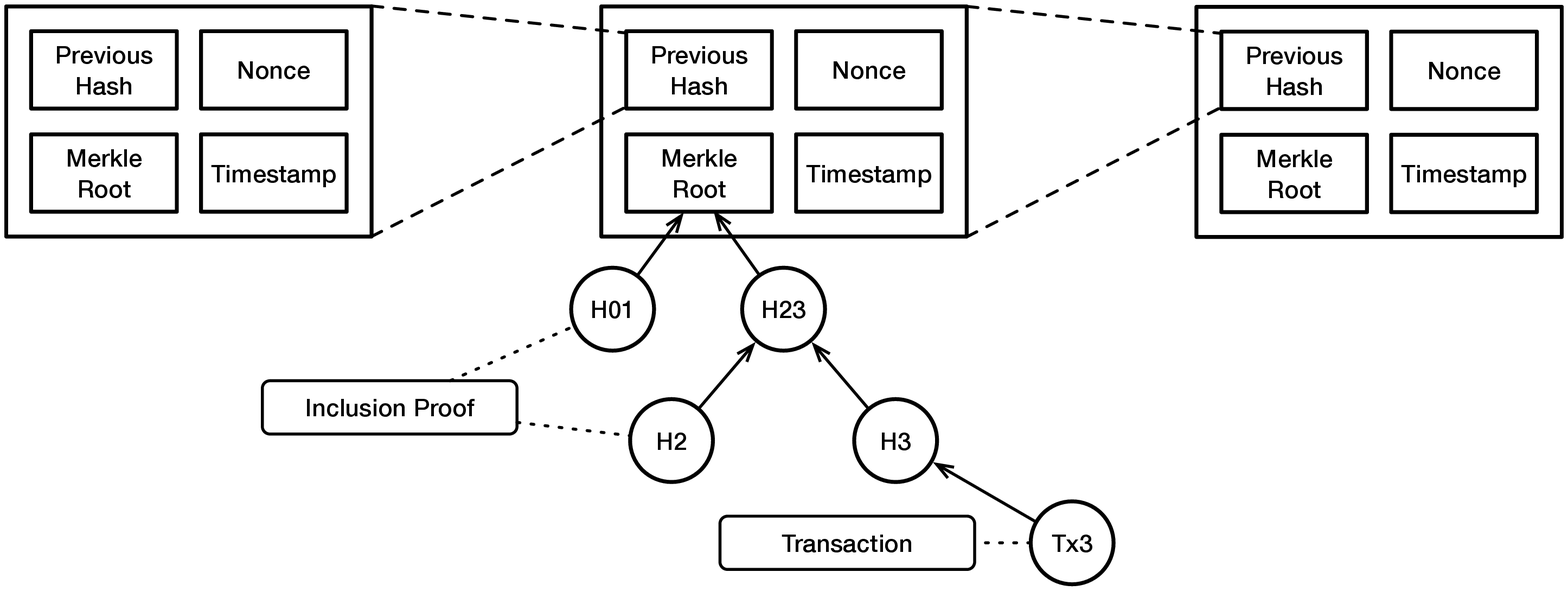}
	\caption{Transactions are included into block headers using Merkle trees.
	In this simplified example, one can prove to someone who holds the root of
	the Merkle tree that \textit{Tx3} is included in the corresponding
	block by showing hashes \textit{H2} and \textit{H01}. In \name, this
	mechanism is used to prove to clients that certificates are logged in the
	blockchain.}
    \label{block_light}
\end{figure}

As described in Section~\ref{background:blockchain}, light clients keep all
block headers, and each header contains, among other information, the root of
the
Merkle tree containing all transactions of the block. To check whether the
transaction is present in a block, the client reads the Merkle
root from the block and checks whether the inclusion proof delivered with the
certificate is correct. Figure~\ref{block_light} illustrates block headers with
the inclusion proof of a transaction.
Light clients could get the inclusion proof from peers using a dedicated
protocol, but asking for such a
proof would leak information about visited domains. Consequently, to preserve
user privacy, the inclusion proof must be delivered by the domain.

If all verifications are successful, the client accepts the certificate and
establishes a secure communication channel with the domain. Otherwise, the
connection is dropped and the fraudulent certificate constitutes evidence of
CA misbehavior.

\subsection{Certificate Renewal}

To update a \name certificate, a requester could create a new domain contract
every time, but she would be charged for each contract creation. \name allows
requesters to renew a certificate using the same domain contract that was used
to create it. The requester just transfers funds to the domain contract. The
certificate validity period is automatically updated and the contract is ready
to receive new signatures.  In this way, not only the number of contracts with
expired certificates running on the blockchain is minimized, but the cost of
renewing a certificate is also reduced.

\section{Security Analysis}
\label{sec:security}

The adversary can create a fraudulent certificate if and only if (a) it can
compromise $i$ CAs, (b) it can conduct a successful domain validation with $j$
honest CAs (e.g., by attacking the CA's DNS resolution process, or by launching
a MITM attack on a link between the CA and the domain server), and (c) $i+j \ge
T$. We consider that such an attack requires a tremendous effort from the
attacker and goes beyond the scope of our adversary model.
Nevertheless, even if the adversary succeeds, \name allows detection of such an
attack with high probability. Clients, while verifying the certificate, have an
inclusion proof of the corresponding transaction in a block. Clients keeping
block headers locally can directly check the proof.

We tested our three Ethereum contracts for security vulnerabilities using the
\emph{Securify} scanner by Tsankov et al.~\cite{tsankov2018securify}. This
allowed us to verify that an attacker cannot freeze or steal Ether, as it
happened in July 2017 with Parity's multi-signature wallet, for example.

\name does not have any single
point of failure.  Undermining the availability of \name to a requester or a CA
running a blockchain node requires an adversary to block all connections to
other peer nodes. To block \name globally, one needs to launch a 51\% attack.
If the adversary controls a significant portion of the network's computational
power but not the majority of it, she can delay the issuance of certificates,
but eventually all blockchain transactions important for \name operation will
be included in the blockchain by honest miners.

An authorized CA could misbehave by sending some meaningless data as a signature
and receiving a payment.  However, such a strategy would not pay off in the long
term. First, each contract specifies which entities are allowed to send a
signature and obtain money in return. If a CA issues invalid signatures, it will
earn a bad reputation and will be avoided in subsequent certificate issuances.
Note that each interaction with the contract will be committed to the blockchain
so any dishonest behavior of a CA becomes permanently visible.  Second, it is
possible to prevent such misbehavior by checking the validity of the signature
in the domain contract itself. However, this option is not the default approach
in our system, since performing computationally intensive computations, such as
public-key cryptography, in a smart contract running on a blockchain entails
higher usage of nodes' computational power and consequently higher costs for the
domain owner to obtain the certificate.

An authorized CA could also simply ignore a certificate issuance request in
which it is listed. This situation is easily mitigated either by listing a
number of authorized CAs greater than $T$ in the domain contract, as we discuss
in Section~\ref{sec:discussion}, or by canceling the unsuccessful request and
creating a new one that does not list the blocking CA. Again, a misbehaving
CA would earn a bad reputation by remaining unresponsive to issuance requests.

Schnorr multi-signatures are known to be vulnerable to rogue-key
attacks~\cite{bellare2006multi}. A malicious CA could choose a key as a function
of that of other CAs in such a way that it can then forge a multi-signature.
This attack is prevented by requiring signers to prove knowledge of their own
secret key. That property is respected in \name as CA certificates are
self-signed and known in advance by clients.

\section{Realization in Practice}
\label{sec:implementation}

Although our implementation of \name is based on the Ethereum blockchain, most
of the concepts we have presented so far are blockchain agnostic and could be
implemented on any other decentralized platform that supports sufficiently
expressive smart contracts. Ethereum~\cite{wood2014ethereum} is a
blockchain-based platform mainly aimed towards running smart contracts. Since
its launch in July 2015, it became the second most popular cryptocurrency with a
total market capitalization of more than \$20B, as of September
2018~\cite{CoinMarketCap}. Transaction cost in Ethereum depends on the amount of
computation performed by the smart contract (triggered by a transaction). Each
operation (e.g., addition or storing a byte in a contract) has a constant cost
expressed in \textit{gas} units~\cite{wood2014ethereum}.  However, the gas cost
expressed in the Ethereum's currency (i.e., \textit{ether}) varies. A
transaction sender specifies how much she is willing to pay for the gas used.
The price implicitly determines the priority of the transaction, as miners are
compensated with the cost of used gas in a block they have mined.

We implemented all contracts (i.e., the central, domain, and storage contracts)
in \textit{Solidity} (version 0.4.7), a Turing-complete programming language
whose syntax is similar to that of JavaScript. Certificates are encoded in JSON
format (an example is presented in Listing~\ref{certificate}). To implement the
Schnorr signature scheme, we used Bitcoin's C elliptic curve
library~\cite{Bitcoin_secp256k1} together with its Python
bindings~\cite{Python_secp256k1}.  We used \textit{secp256k1} as the default
elliptic curve for our implementation. As a hash function for Schnorr signatures
we used SHA256; the Ethereum blockchain uses Keccak-256. Nonces required by the
Schnorr algorithm were generated deterministically as described in
RFC~6979~\cite{rfc6979}.

The requester implementation consists of two modules:
\begin{enumerate*}[label=(\alph*)]
    \item a Go implementation of an Ethereum full node (\textit{Geth}, version
    1.5.5~\cite{geth}), and
    \item a Python application that communicates with the local node (using the
    \textit{Web3.py} library~\cite{web3py}).
\end{enumerate*}
The latter module is used to automate the process of requesting and creating
certificates.  The application takes certificate parameters and creates the
transaction calling the \texttt{create\-Domain\-Contract} method of the central
contract (see Listing~\ref{centralContract}).  After the domain contract is
created, the application observes it and listens for the
\texttt{all\-Cert\-Signatures\-Gathered} event, which informs that co-signing of
the certificate is finished (see Listing~\ref{domainContract}). When a block
with this event appears, the application reads all partial signatures from the
contract and merges them. Then, the transaction with certificate data is sent to
the storage contract. To ensure that the transaction is not overridden,
the requester waits for a few confirmation blocks before obtaining the proof
of inclusion. In the current deployment of Ethereum, 12 confirmation blocks is
a commonly accepted standard~\cite{vitalikForks}.

CAs, similarly to requesters, run Ethereum nodes using Geth, and implement a
Python application to communicate with it. CAs observe the central contract and
wait for the \texttt{new\-Domain\-Contract} event. When the event is triggered,
each CA checks whether it is listed as an authorized CA. If so, the CA performs
domain validation, as specified by the ACME specification (see
Section~\ref{acme}). If the verification succeeds, the CA deterministically
generates a private-public nonce pair from certificate data encoded in the JSON
format with alphabetically sorted keys. Then, it submits the public nonce to the
domain contract and waits for the \texttt{allCertNoncesGathered} event. Once the
event is received, the CA reads the nonces of the other authorized CAs and sends
its partial Schnorr signature to the domain contract.

\section{Evaluation}
\label{sec:evaluation}

We evaluated \name on the \textit{Ropsten} test network, which is
the default Ethereum development environment. This test network
provides the same functionalities and characteristics as the main network,
but the ether on Ropsten has no monetary value.
To conduct experiments, we created a setting with up to 20 CAs, a requester (and
its servers), and a client. CAs and the requester are running the scripts
described in Section~\ref{sec:implementation}. The client software was run on
a commodity machine with an Intel Core i5-2410M 2.30GHz processor, 4GB of RAM,
and Ubuntu 16.04.

\subsection{Time Required for Issuing a Certificate}

\begin{figure}
    \centering
    \begin{subfigure}[b]{0.4\columnwidth}
        \centering
        \includegraphics[width=\columnwidth]{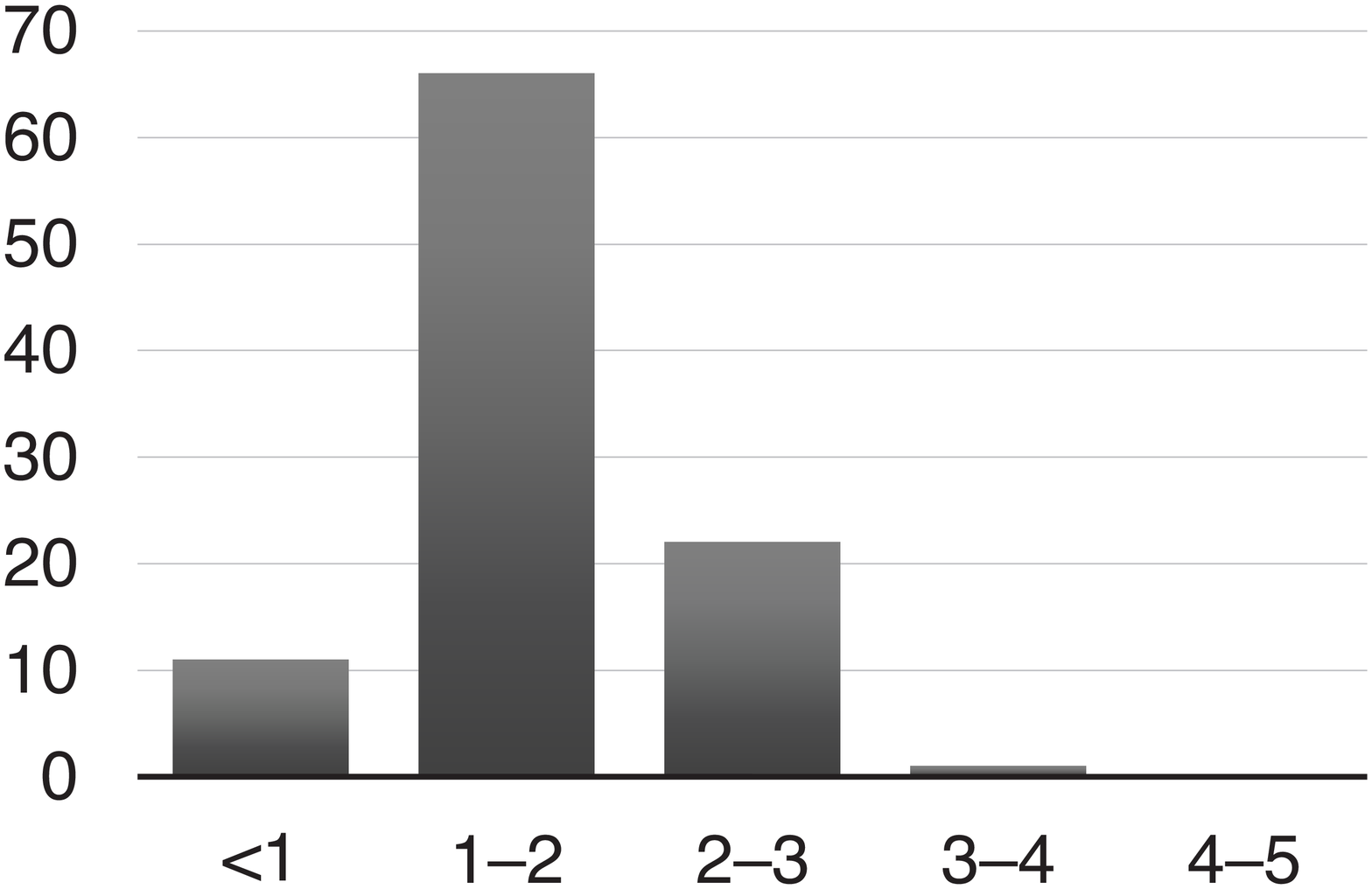}
        \caption{$T = 2$}
    \end{subfigure}
    \hspace{2em}
    \begin{subfigure}[b]{0.4\columnwidth}
        \centering
        \includegraphics[width=\columnwidth]{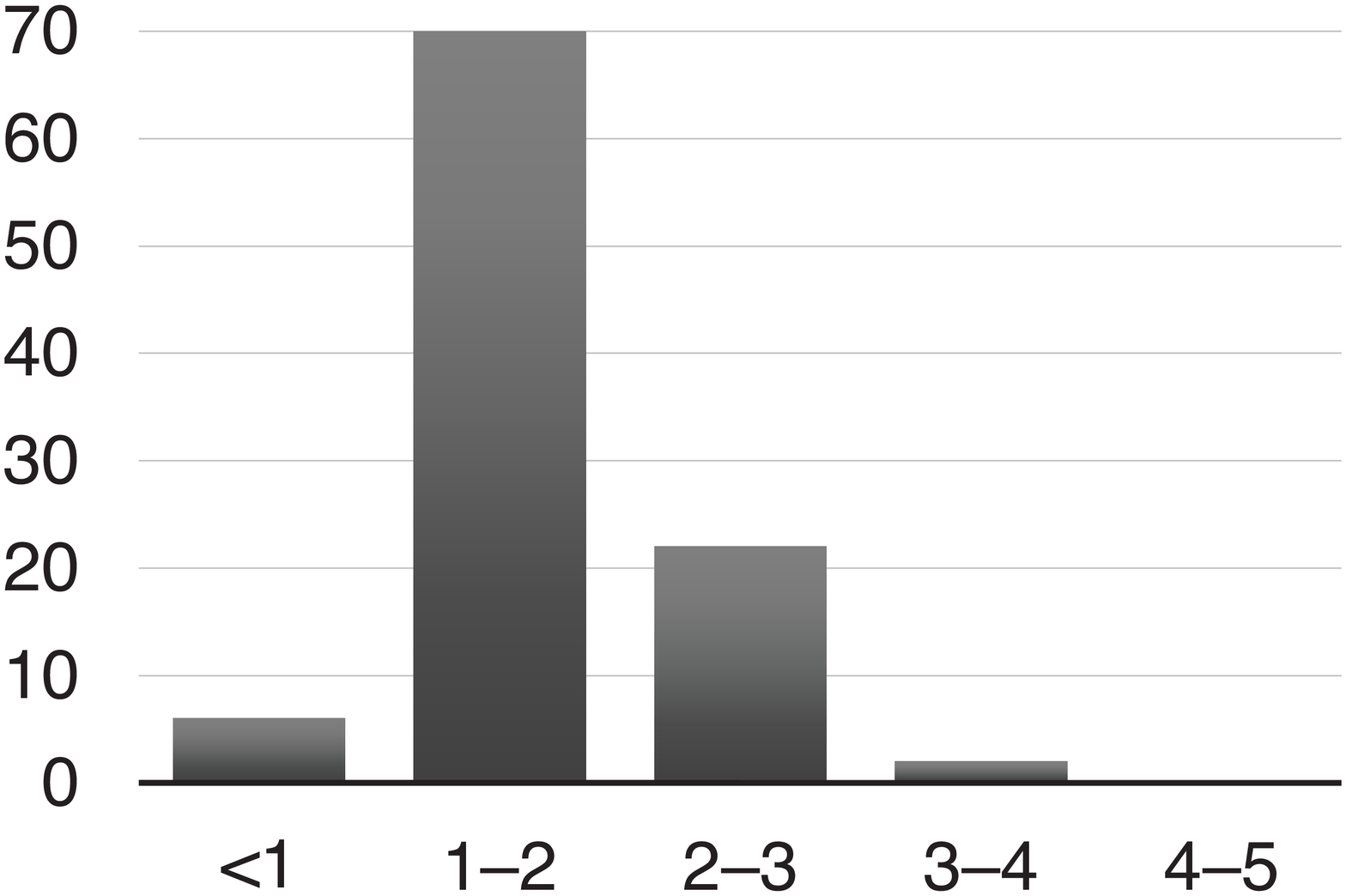}
        \caption{$T = 5$}
    \end{subfigure}
    
    \vspace{1ex}
    \begin{subfigure}[b]{0.4\columnwidth}
        \centering
        \includegraphics[width=\columnwidth]{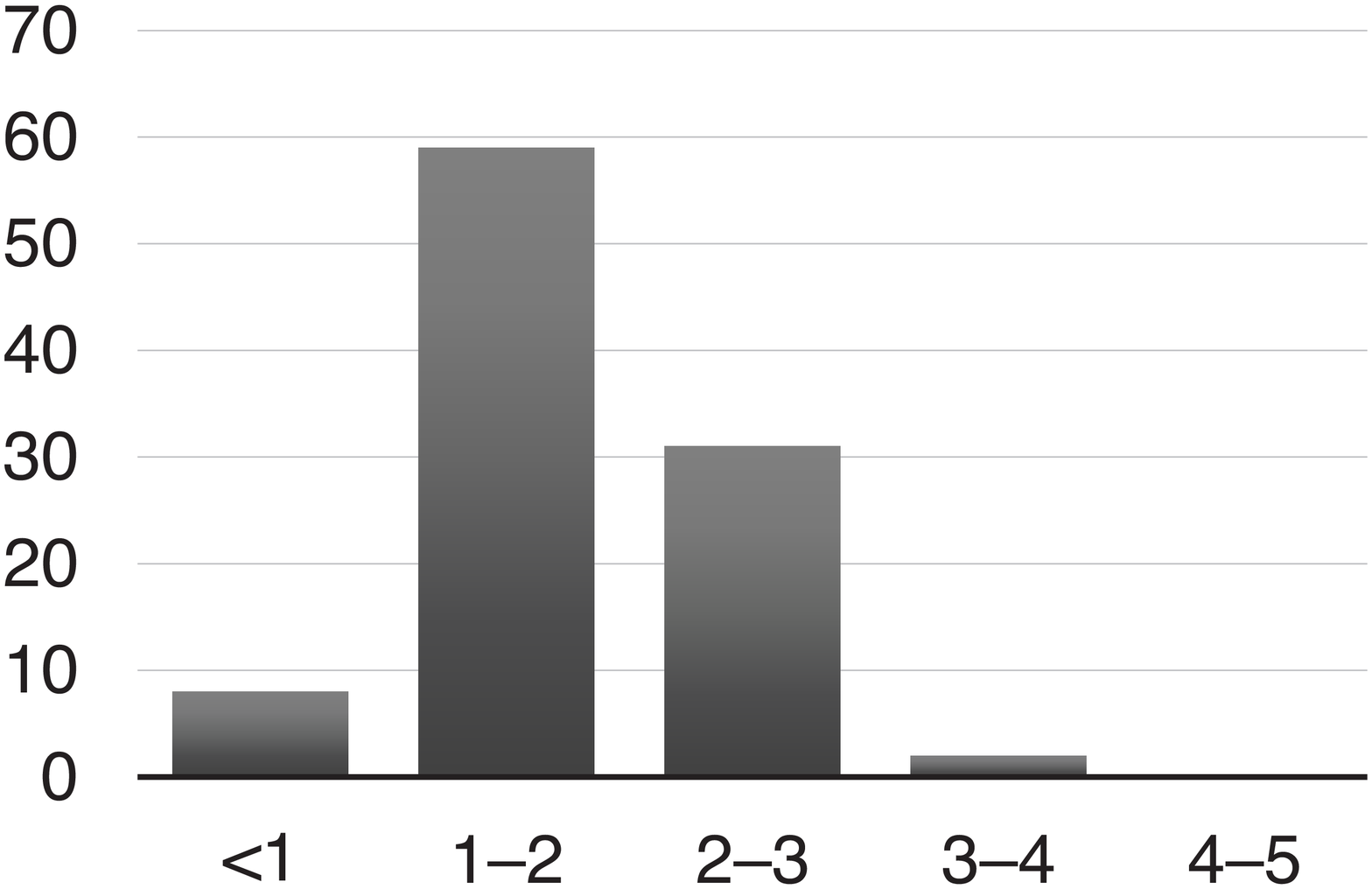}
        \caption{$T = 10$}
    \end{subfigure}
    \hspace{2em}
    \begin{subfigure}[b]{0.4\columnwidth}
        \centering
        \includegraphics[width=\columnwidth]{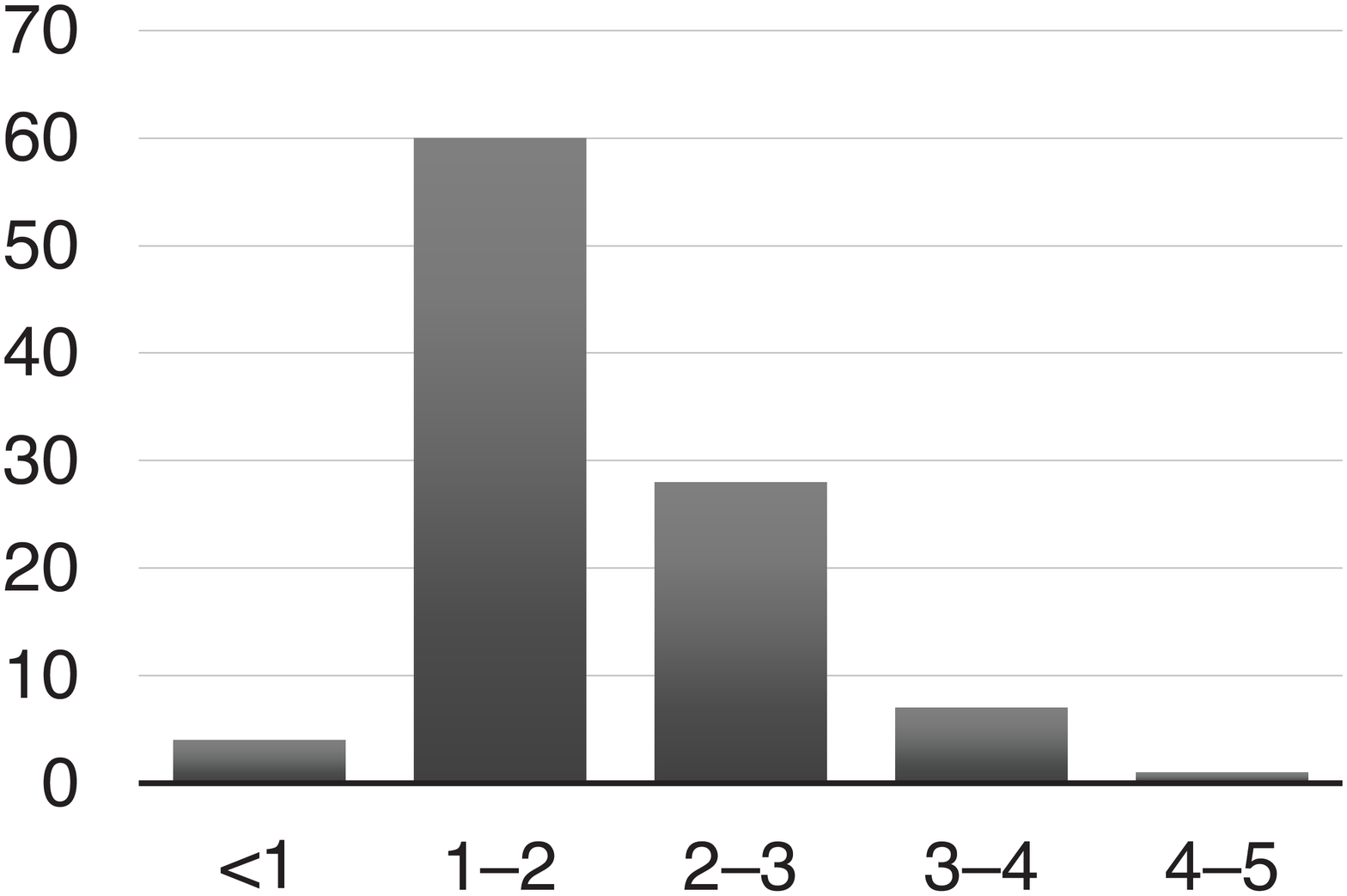}
        \caption{$T = 20$}
    \end{subfigure}
    \vspace{1ex}
    \caption{Certificate issuance time distribution (minutes on the x-axis,
    \% on the y-axis), for different thresholds.}
    \label{issuance_duration}
\end{figure}

First, we examine the time needed to obtain the certificate using \name, and how
the threshold number $T$ of the required CAs affects certificate creation time.
To highlight the actual effect of parameter $T$, we conducted measurements in a
setting where the requester obtains the inclusion proof just after a block with
his transaction is mined (without waiting for any block confirmations).
The obtained results are presented in Figure~\ref{issuance_duration}. The
duration,
in most cases, is less than two minutes and varies only slightly with
the number of required CAs. When taking into account the delay caused by waiting
for 12 blocks, the average issuance duration would increase by three
minutes, since block creation in Ethereum takes around 15 seconds on
average~\cite{etherscan}.

The required time is greatly influenced by the frequency of blockchain updates.
The process of gathering signatures requires at least four blocks.  In
the best-case scenario, the first block already contains the transaction that
creates a new domain contract, so that all CAs can notice a pending certificate
request. Gathering signatures involves two rounds, one for gathering nonces and
one for obtaining partial signatures. Each round
can start only when the previous round is over. Assuming that domain
verification is performed before the next block is found, CAs may broadcast
their transactions with the public nonces and have the transactions
included in the second block. Once all nonces are embedded in the
blockchain, CAs may put their partial signatures in the third block. Then,
one additional block is needed to put the requester's transaction with
certificate data into the blockchain.

As our results show, the broadcast transaction is rarely put in the very next
block. Usually, it is included in the second or the third one, because network
latency plays an important role in blockchains with small block times. For
comparison, Bitcoin blocks are mined every 10 minutes~\cite{BTCblocktime}.

\subsection{Cost of Issuing a Certificate}

\begin{figure}
    \centering
    \includegraphics[width=.85\columnwidth]{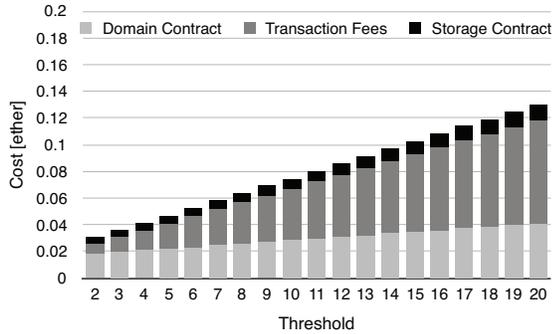}
    \caption{Certificate creation cost, in function of the threshold $T$,
    including the creation of the domain contract, the cumulated fees
    of all transactions sent from CAs and the transaction to the storage
    contract.}
    \label{num_cas_cost}
\end{figure}

We now investigate the cost of using \name in Ethereum.  For the evaluation, we
set the cost of gas to 20 Gwei, which is above the average price for gas in
Ethereum~\cite{etherscan}. Figure~\ref{num_cas_cost} shows the linear increase
of cost in function of the number of CAs involved in creating a certificate. The
cost we measure is the total fee of all transactions sent during creation of the
certificate (without compensations). The first---and most
expensive---transaction is initiated by the requester and results in the
creation of a domain contract. The contract code remains the same for each
request, but the content of its variables may differ. For the evaluation we used
a 32-byte domain public key and 14-byte domain name. The next transactions are
sent by CAs; each CA sends two transactions, so in total $2T$ transactions are
broadcast by the CAs. The last transaction is sent by the requester to the
storage contract.

As domain contracts are reused for certificate renewal, requesters do
not pay again for the creation of the domain contract, but only for
transaction fees (see Figure~\ref{num_cas_cost}).
Apart from domain contracts created for each certificate request, the central
contract and the storage contract are also present in the blockchain. The
one-time cost of creating the central contract and the storage contract is
negligible. Their creators do not have any privileges in \name.

We conduct our cost evaluation in a conservative setting, but some
modifications could lower the costs. For example, we could move the system
to Ethereum Classic, an alternative version of the Ethereum blockchain.
At the time of writing, the price of ether on Ethereum Classic oscillates
around 11 USD, which is roughly 20 times lower than the price on Ethereum.
The price of gas in ether is roughly equal on both blockchains, thus we
could decrease the costs by a factor of 20. However, Ethereum Classic has
a slower hash rate than that of Ethereum~\cite{ETCexplorer},
which means that launching a 51\% attack is proportionally easier.

\subsection{Time Required for Verifying a Certificate}

Another aspect we investigate is the computation overhead on the client side. We
again examine this factor with respect to the threshold number $T$
of CAs. The measurements were performed on the commodity computer described
earlier. Each measurement was repeated 100 times and the average value of all
100 samples was taken as a result.

\begin{table}
	\footnotesize
	\centering
	\begin{tabular}{@{}l @{}c@{}c@{} c @{}c@{}c@{} c @{}c@{}c@{} c @{}c@{}c@{}}
		\toprule
		
		& \multicolumn{2}{c}{\textbf{2 CAs}} &&
		  \multicolumn{2}{c}{\textbf{5 CAs}} &&
		  \multicolumn{2}{c}{\textbf{10 CAs}} &&
		  \multicolumn{2}{c}{\textbf{20 CAs}} \\
		  
		\cmidrule{2-3} \cmidrule{5-6} \cmidrule{8-9} \cmidrule{11-12}
		
		& \multicolumn{1}{c}{[ms]} & \multicolumn{1}{c}{\%} &&
		  \multicolumn{1}{c}{[ms]} & \multicolumn{1}{c}{\%} &&
		  \multicolumn{1}{c}{[ms]} & \multicolumn{1}{c}{\%} &&
		  \multicolumn{1}{c}{[ms]} & \multicolumn{1}{c}{\%} \\
		  
		\midrule
		
		\textbf{key combination} & 0.036 & 17 && 0.042 & 19 && 0.046 & 20 && 0.052 & 24 \\
		\textbf{block inclusion} & 0.073 & 34 && 0.069 & 32 && 0.078 & 33 && 0.067 & 30 \\
		\textbf{signature verif.} & 0.105 & 49 && 0.108 & 49 && 0.109 & 47 && 0.102 & 46 \\
		
		\bottomrule
	\end{tabular}
	\smallskip
	\caption{Duration of certificate verification steps.}
	\label{duration_table}
\end{table}

As Table~\ref{duration_table} shows, the signature verification time is little
influenced by $T$. Thanks to the Schnorr multi-signature, the signature
verification
with a combined key does not differ from the standard verification of a single
Schnorr signature. The time needed to combine the keys increases as the number
of keys increases. Nevertheless, key combination accounts for 17\% of the total
time when $T = 2$, and for 24\% when $T = 20$, so the difference is unnoticeable
from the user's perspective. Verifying the inclusion proof takes about one third
of the total time.
Ethereum uses Patricia trees~\cite{PatriciaTree} (a variant of Merkle
trees) for including transactions in blocks. For the evaluation
we used a block containing 10 other transactions, a commonly observed number
of transactions in a block on the Ethereum main network~\cite{etherscan}.
Finally, the verification of the merged signature proceeds in constant time and
takes about 0.1~ms.

\section{Discussion}
\label{sec:discussion}

For the sake of simplicity, we assumed that certificates are always signed by
$T$ CAs.  However, it would be possible for a requester to list in
his domain contract a number of authorized CAs greater than $T$. The contract
would ensure that only $T$ CAs contribute to the multi-signature by accepting
the first $T$ nonces and disallowing CAs that did not include a nonce to send
signatures.
More authorized CAs implies a possibly faster issuance of signatures as only
the first $T$ CAs would be able to co-sign the certificate and receive financial
compensations, thus CAs would have a strong incentive to conduct the domain
verification process as fast as possible. On the other hand, if the requester
includes many CAs, then the cost of creating such a contract becomes higher (as
the requester must pay for the additional storage needed to hold the CAs'
blockchain addresses and the table with their compensations).
As we expect \name-compatible CAs to respond to valid certificate issuance
requests in a large majority of cases (since they are financially rewarded for
doing so), it is probably best for requesters to list exactly $T$ accepted CAs,
unless experience proves it is desirable to do otherwise.
It is also possible for a requester to use a value of $T$ greater than the
value commonly used by browsers.
Such a change would make a certificate more resilient, as revoking
one CA would not invalidate all the certificates it has issued.
Such a high-resilience, high-security certificate would naturally be more
expensive, however.

Root CAs, which are directly trusted by the browser or OS, rarely sign domain
certificates directly. Instead, they sign intermediate CAs, which in turn issue
domain certificates. Although, for presentation purposes, we simplified the
usual CA hierarchy, \name supports longer certificate chains (i.e., with
intermediate CAs). To request a certificate, the requester does not have to
specify a particular intermediate CA. It is sufficient to include the blockchain
address of a root CA in the domain contract. Thereafter, all intermediate CAs
can use the same blockchain account to send their partial signatures. If a
certificate in \name is signed by intermediate CAs, then the domain owner must
provide its clients with the chain of trust up to the root CAs, as well as
proofs of key possession (to prevent rogue-key attacks). Browsers first verify
whether the intermediate CAs are trusted and then verify the certificate (as
described in Section~\ref{section_certificate_verification}) using the public
keys of the intermediate CAs that signed the certificate.

By automating all operations, we facilitate the usage of short-lived
certificates, which reduce the attack window in the event where
a certificate would be illegitimately issued.
Although it is hard to determine what the optimal certificate validity
period would be in practice, we envision \name certificates to have a
lifetime of around 90 days initially, as is currently the case with Let's
Encrypt.
Short-lived, multi-signed certificates greatly reduce the need for a
revocation system, but do not completely suppress it. Given that designing
a satisfactory revocation system has proven to be an extremely challenging
task~\cite{liu2015emc,PKISN,RITM2016}, we consider it to go beyond the scope
of this paper. Nevertheless, the security of \name can be further improved
by combining it with any existing revocation scheme.

\section{Related Work}
\label{sec:related}

Several existing systems rely on blockchains to provide PKI functionalities. For
instance, Namecoin~\cite{loibl2014namecoin} allows users to register a name and
attach data, such as a public-key fingerprint, to that name. Namecoin has its
own dedicated namespace (\texttt{.bit} top-level domain), which is resolved
using the Namecoin infrastructure (without involving DNS).
CertCoin~\cite{fromknecht2014certcoin} is an improvement over Namecoin, which
reduces the amount of storage necessary to use the system, and introduces key
revocation and recovery functionalities. In Ethereum, the resolution of
human-readable names into identifiers of digital resources (for example
blockchain addresses of smart contracts)  is provided by the Ethereum Name
Service~\cite{ens}. In contrast to previous systems, such as Namecoin or
CertCoin, \name allows domain owners to obtain certificates for standard DNS
names.

BKI~\cite{wan2017bki} is a blockchain-based PKI that involves an adjustable
number of CAs to issue certificates, but it does not make use of
multi-signatures to make certificate verification more scalable (as \name does).
Moreover, BKI requires that all clients contact a third party (a
``blockchain-based log maintainer'') during certificate verification, which is
problematic for latency and privacy reasons. \name also improves upon BKI by
allowing CAs to get remunerated automatically through smart contracts.
SCPKI~\cite{al2017scpki} is another blockchain-based PKI, but it departs from
the CA model used by TLS and relies instead on a web-of-trust model to solve a
range of identity-related problems. Doing away with CAs entirely to instead rely
upon individuals to certify the authenticity of public keys is a model that may
be appropriate for user-to-user interactions, but it is not adapted to today's
web. Therefore, the assumptions and objectives underlying SCPKI are vastly
different from those of \name.

As an enhancement to the standard web PKI, Google developed Certificate
Transparency (CT)~\cite{rfc6962} to detect targeted attacks by making all the
certificates submitted to their log servers publicly visible. To work
effectively, CT requires that every certificate is accompanied by a signed
statement that a log has received the certificate (and will add it to its
store). Logs in CT must be highly available as they are necessary for issuing
certificates. Moreover, CT does not provide any attack prevention as it is
designed exclusively for attack detection. AKI~\cite{kim2013aki} and its
successor ARPKI~\cite{basin2014aar} rely on certificate logs similar to CT's,
but provide additional security guarantees: they employ multi-signature
certificates signed by $n$ CAs, and aim at protecting users from an adversary
who would be able to compromise up to $n - 1$ CAs. These systems rely on  log
servers whose availability is crucial for standard operations. Moreover, AKI and
ARPKI lack an automated framework for requests and payment.

The monitoring of log systems is another important aspect investigated by
previous work. Chuat et al.~\cite{chuat2015efficient} and Nordberg et
al.~\cite{nordberg2015gossiping} have proposed systems for TLS clients to gossip
about certificate logs (using regular web traffic). Leveraging a blockchain to
monitor public logs was proposed by Bonneau in EthIKS~\cite{bonneau2016ethiks}.
The system's objective is to enhance auditability of the CONIKS
system~\cite{melara2015coniks}, which is a log-based end-user key verification
service.  In CONIKS, users are required to monitor the correctness of their own
data in the repository or trust a third party to perform the audit on their
behalf.  EthIKS, by incorporating CONIKS data structures in a smart contract and
relying on the Ethereum network to enforce honest handling of the repository,
reduces the trust put on other users or third-party auditors.
IKP~\cite{Reischuk16:IKP} introduces a blockchain-based system that provides
financial incentives for detecting fraudulent certificates. The system leverages
smart contracts, allows domain owners to specify their trusted CAs, and allows
CAs to create an escrowed insurance fund to protect against CA misbehavior.
Anyone who can find a certificate non-compliant with a domain policy can be
compensated by the CA's insurance.

Catena~\cite{tomescucatena} proposes a blockchain implementation of a log system
that is accessible to lightweight blockchain clients. It leverages Bitcoin's
double-spending prevention mechanism. Each Catena log statement is put into a
Bitcoin transaction. Clients verify the existence of transactions using a light
client protocol in Bitcoin (called Simplified Payment Verification).
Consequently, the equivocation of the log is equivalent to double-spending,
which in turn is as hard as forking the Bitcoin blockchain.

Notaries~\cite{wendlandt2008perspectives,marlinspikeconvergence} can help
detecting MITM attacks with multi-path probing. By contacting notary servers,
clients can verify whether a domain's certificate is also observed from a set of
vantage points. The notary server contacts the domain and forwards its view of
the domain's certificate to the client. Unfortunately, contacting a notary
service introduces significant latency, and the achieved level of security is
hard to assess without any mechanism for detecting malicious notary servers.

CoSi~\cite{syta2016keeping} proposes an efficient way of using multi-signatures
to co-sign statements issued by CAs. Each statement needs to be co-signed by a
threshold number of witnesses in order to be accepted by clients.  Consequently,
even if an attacker compromises an authority, all malicious statements need to
be publicly exposed before they can be used for an attack.  However, CoSi
requires coordination in the co-signing protocol and relies on direct
communication between witnesses. \name does not rely on those assumptions, as it
still works effectively when signers do not communicate directly with each
other. Furthermore, CoSi's security is still only as strong as the weakest link,
as witnesses only approve the statements issued by CAs~\cite{syta2016keeping}
and do not conduct a full domain validation themselves. An attacker could still
exploit vulnerabilities, in BGP for example, to hijack traffic destined to a
victim's domain~\cite{birge18bamboozling}.

In opposition to existing approaches, \name is, to the best of our knowledge, the
first PKI to
\begin{enumerate*}[label=(\alph*)]
	\item require that multiple CAs perform a complete domain validation from
	different vantage points for an increased resilience to compromise and
	hijacking,
	\item scale to a high number of CAs by using an efficient multi-signature
	scheme, and
	\item provide a framework for paying multiple CAs automatically.
\end{enumerate*}

\section{Conclusion}
\label{sec:conclusion}

We observed that blockchains and smart contracts lend themselves to a novel PKI
design that does not rely on any globally trusted entities. Our implementation
in Ethereum demonstrates the viability of the approach. Our system is
conceptually simple, yet achieves several surprising and desirable properties:
fully autonomous operation, the creation of a malicious certificate requires
compromising a large number of trusted entities and still becomes globally
visible, and CAs have a viable business model where they are compensated for
their validations and signatures. We hope that \name constitutes a worthwhile
contribution in the quest towards a highly secure and usable PKI.

\section*{Acknowledgments}

The research leading to these results has received funding from the
European Research Council under the European Union's Seventh Framework
Programme (FP7/2007-2013) / ERC grant agreement 617605. We gratefully
acknowledge support from ETH Zurich and from the Zurich Information
Security and Privacy Center (ZISC).
Pawel's work was supported by the SUTD SRG ISTD 2017 128 grant.

\footnotesize{
\bibliographystyle{abbrv}
\bibliography{references}

\begin{thebibliography}{10}

\bibitem{letsencrypt_cost}
J.~Aas.
\newblock Launching our crowdfunding campaign.
\newblock \url{https://perma.cc/5DXM-PMY8}, November 2016.

\bibitem{al2017scpki}
M.~Al-Bassam.
\newblock {SCPKI}: a smart contract-based {PKI} and identity system.
\newblock In {\em Proceedings of the ACM Workshop on Blockchain,
  Cryptocurrencies, and Contracts}. ACM, 2017.

\bibitem{barnes2015automatic}
R.~Barnes, J.~Hoffman-Andrews, and J.~Kasten.
\newblock Automatic certificate management environment ({ACME}).
\newblock {IETF} draft, August 2018.

\bibitem{basin2014aar}
D.~Basin, C.~Cremers, T.~H.-J. Kim, A.~Perrig, R.~Sasse, and P.~Szalachowski.
\newblock {ARPKI: Attack Resilient Public-Key Infrastructure}.
\newblock In {\em Proceedings of the ACM Conference on Computer and
  Communications Security (CCS)}, 2014.

\bibitem{bellare2006multi}
M.~Bellare and G.~Neven.
\newblock Multi-signatures in the plain public-key model and a general forking
  lemma.
\newblock In {\em Proceedings of the ACM conference on Computer and
  Communications Security (CCS)}, 2006.

\bibitem{birge18bamboozling}
H.~Birge-Lee, Y.~Sun, A.~Edmundson, J.~Rexford, and P.~Mittal.
\newblock Bamboozling certificate authorities with {BGP}.
\newblock In {\em Proceedings of the USENIX Security Symposium}, 2018.

\bibitem{bonneau2016ethiks}
J.~Bonneau.
\newblock {EthIKS}: Using {Ethereum} to audit a {CONIKS} key transparency log.
\newblock In {\em Proceedings of the International Conference on Financial
  Cryptography and Data Security}, 2016.

\bibitem{PatriciaTree}
E.~Buchman.
\newblock Understanding the {Ethereum} trie.
\newblock \url{https://perma.cc/22C3-BAMB}, June 2014.

\bibitem{vitalikForks}
V.~Buterin.
\newblock On slow and fast block times.
\newblock \url{https://perma.cc/8556-WSRV}, September 2015.

\bibitem{chuat2015efficient}
L.~Chuat, P.~Szalachowski, A.~Perrig, B.~Laurie, and E.~Messeri.
\newblock Efficient gossip protocols for verifying the consistency of
  certificate logs.
\newblock In {\em Proceeding of the IEEE Conference on Communications and
  Network Security (CNS)}, 2015.

\bibitem{CoinMarketCap}
{CoinMarketCap}.
\newblock Cryptocurrency market capitalizations.
\newblock \url{https://coinmarketcap.com}.

\bibitem{geth}
{Ethereum}.
\newblock Official {Go} implementation of the {Ethereum} protocol.
\newblock https://geth.ethereum.org/.

\bibitem{etherscan}
Etherscan.io.
\newblock Ethereum charts \& statistics.
\newblock \url{https://etherscan.io/charts}.

\bibitem{felt2016rethinking}
A.~P. Felt, R.~W. Reeder, A.~Ainslie, H.~Harris, M.~Walker, C.~Thompson, M.~E.
  Acer, E.~Morant, and S.~Consolvo.
\newblock Rethinking connection security indicators.
\newblock In {\em Proceedings of the Symposium on Usable Privacy and Security
  (SOUPS)}, 2016.

\bibitem{Fisher2015Apple}
D.~Fisher.
\newblock Apple pushing developer toward {HTTPS} connections from apps.
\newblock https://perma.cc/6QGZ-AQAB, June 2015.

\bibitem{fromknecht2014certcoin}
C.~Fromknecht, D.~Velicanu, and S.~Yakoubov.
\newblock A decentralized public key infrastructure with identity retention.
\newblock {\em IACR preprint}, 2014.

\bibitem{ETCexplorer}
Gastracker.io.
\newblock Ethereum classic block explorer.
\newblock \url{https://gastracker.io/}.

\bibitem{Bitcoin_secp256k1}
GitHub.
\newblock Bitcoin secp256k1 curve library.
\newblock \url{https://github.com/bitcoin-core/secp256k1}.

\bibitem{Python_secp256k1}
GitHub.
\newblock {Python} {FFI} bindings for secp256k1.
\newblock \url{https://github.com/ludbb/secp256k1-py}.

\bibitem{web3py}
GitHub.
\newblock {Web3.py} library.
\newblock \url{https://github.com/pipermerriam/web3.py}.

\bibitem{ens}
N.~Johnson.
\newblock {Ethereum} domain name service.
\newblock EIP 137, April 2016.

\bibitem{kim2013aki}
T.~H.-J. Kim, L.-S. Huang, A.~Perrig, C.~Jackson, and V.~Gligor.
\newblock Accountable key infrastructure ({AKI}): A proposal for a public-key
  validation infrastructure.
\newblock In {\em Proceedings of the ACM International Conference on World Wide
  Web (WWW)}, 2013.

\bibitem{rfc6962}
B.~Laurie, A.~Langley, and E.~Kasper.
\newblock Certificate transparency.
\newblock RFC 6962, June 2013.

\bibitem{LetsEncryptStats}
{Let's Encrypt}.
\newblock Statistics.
\newblock \url{https://letsencrypt.org/stats/}.

\bibitem{liu2015emc}
Y.~Liu, W.~Tome, L.~Zhang, D.~Choffnes, D.~Levin, B.~Maggs, A.~Mislove,
  A.~Schulman, and C.~Wilson.
\newblock An end-to-end measurement of certificate revocation in the web's
  {PKI}.
\newblock In {\em Proceedings of the ACM Internet Measurements Conference
  (IMC)}, 2015.

\bibitem{marlinspikeconvergence}
M.~Marlinspike.
\newblock {SSL} and the future of authenticity, 2011.

\bibitem{Reischuk16:IKP}
S.~Matsumoto and R.~M. Reischuk.
\newblock {IKP}: Turning a {PKI} around with blockchains.
\newblock In {\em Proceedings of the IEEE Symposium on Security and Privacy
  (S\&P)}, 2017.

\bibitem{melara2015coniks}
M.~S. Melara, A.~Blankstein, J.~Bonneau, E.~W. Felten, and M.~J. Freedman.
\newblock {CONIKS}: Bringing key transparency to end users.
\newblock In {\em Proceedings of the USENIX Security Symposium}, 2015.

\bibitem{merzdovnik2016whom}
G.~Merzdovnik, K.~Falb, M.~Schmiedecker, A.~G. Voyiatzis, and E.~Weippl.
\newblock Whom you gonna trust? {A} longitudinal study on {TLS} notary
  services.
\newblock In {\em Proceedings of the IFIP Annual Conference on Data and
  Applications Security and Privacy}, 2016.

\bibitem{loibl2014namecoin}
Namecoin.
\newblock \url{https://namecoin.org/}.

\bibitem{nordberg2015gossiping}
L.~Nordberg, D.~K. Gillmor, and T.~Ritter.
\newblock Gossiping in {CT}.
\newblock Internet draft, January 2018.

\bibitem{okamoto1999multi}
K.~Ohta and T.~Okamoto.
\newblock Multi-signature schemes secure against active insider attacks.
\newblock {\em IEICE Transactions on Fundamentals of Electronics,
  Communications and Computer Sciences}, 82(1):21--31, 1999.

\bibitem{rfc6979}
T.~Pornin.
\newblock Deterministic usage of the digital signature algorithm ({DSA}) and
  elliptic curve digital signature algorithm ({ECDSA}).
\newblock RFC 6979, 2013.

\bibitem{ChromeIndicator2016}
E.~Schechter.
\newblock Moving towards a more secure web.
\newblock \url{https://perma.cc/598B-KHQ4}, September 2016.

\bibitem{schnorr1989efficient}
C.-P. Schnorr.
\newblock Efficient identification and signatures for smart cards.
\newblock In {\em Conference on the Theory and Application of Cryptology},
  1989.

\bibitem{BTCblocktime}
Smartbit.
\newblock Time between blocks.
\newblock \url{https://www.smartbit.com.au/charts/block-interval}.

\bibitem{syta2016keeping}
E.~Syta, I.~Tamas, D.~Visher, D.~I. Wolinsky, P.~Jovanovic, L.~Gasser,
  N.~Gailly, I.~Khoffi, and B.~Ford.
\newblock Keeping authorities ``honest or bust'' with decentralized witness
  cosigning.
\newblock In {\em Proceedings of the IEEE Symposium on Security and Privacy
  (S\&P)}, 2016.

\bibitem{RITM2016}
P.~Szalachowski, L.~Chuat, T.~Lee, and A.~Perrig.
\newblock {RITM}: Revocation in the middle.
\newblock In {\em Proceedings of the IEEE International Conference on
  Distributed Computing Systems (ICDCS)}, 2016.

\bibitem{PKISN}
P.~Szalachowski, L.~Chuat, and A.~Perrig.
\newblock {PKI} safety net ({PKISN}): Addressing the too-big-to-be-revoked
  problem of the {TLS} ecosystem.
\newblock In {\em Proceedings of the IEEE European Symposium on Security and
  Privacy (Euro S\&P)}, 2016.

\bibitem{tomescucatena}
A.~Tomescu and S.~Devadas.
\newblock Catena: Efficient non-equivocation via {Bitcoin}.
\newblock In {\em Proceedings of the IEEE Symposium on Security and Privacy
  (S\&P)}, 2017.

\bibitem{tsankov2018securify}
P.~Tsankov, A.~Dan, D.~D. Cohen, A.~Gervais, F.~Buenzli, and M.~Vechev.
\newblock Securify: Practical security analysis of smart contracts.
\newblock {\em arXiv preprint}, August 2018.

\bibitem{FirefoxIndicator2016}
T.~Vyas.
\newblock No more passwords over {HTTP}, please!
\newblock \url{https://perma.cc/99MC-VSFY}, January 2016.

\bibitem{wan2017bki}
Z.~Wan, Z.~Guan, F.~Zhuo, and H.~Xian.
\newblock {BKI}: Towards accountable and decentralized public-key
  infrastructure with blockchain.
\newblock In {\em Proceedings of the International Conference on Security and
  Privacy in Communication Networks (SecureComm)}, 2017.

\bibitem{wendlandt2008perspectives}
D.~Wendlandt, D.~G. Andersen, and A.~Perrig.
\newblock Perspectives: Improving {SSH}-style host authentication with
  multi-path probing.
\newblock In {\em Proceedings of the USENIX Annual Technical Conference}, 2008.

\bibitem{wood2014ethereum}
G.~Wood.
\newblock {Ethereum}: A secure decentralised generalised transaction ledger.
\newblock {\em Ethereum Project Yellow Paper}, 2014.

\end{thebibliography}
}

\normalsize
\appendix

\section{Contracts and Certificate Format}

\begin{figure}[h]
 \centering
\begin{minipage}[b]{\columnwidth}
\begin{lstlisting}[style=contract_pseudo, frame=single,
caption=Pseudocode of the central contract., label=centralContract]
contract CentralContract:
  address[] public createdDomainContracts;
  event newDomainContract(address contractAddr);
  
  function createDomainContract(
      _certData,
      _authorizedCAs,
      _compensations):
    if suppliedFunds >= sum(_compensations):
      uint thresholdT = _authorizedCAs.length
      address contractAddr = new DomainContract(
          thresholdT,
          _certData,
          _authorizedCAs,
          _compensations)
      createdDomainContracts.push(contractAddr);
      newDomainContract(contractAddr);
  ...
\end{lstlisting}
\end{minipage}
\end{figure}

\begin{figure}[h]
 \centering
\begin{minipage}[b]{\columnwidth}
\begin{lstlisting}[style=contract_pseudo, frame=single,
caption=Pseudocode of the storage contract., label=storageContract]
contract StorageContract:
  certificate[] public storedCertificates;

  function storeCertificate(
      _certData,
      _CAs,
      _signature):
    storedCertificates.push(
        _certData,
        _CAs,
        _signature)
  ...
\end{lstlisting}
\end{minipage}
\end{figure}

\begin{figure}[h]
\centering
\begin{minipage}[b]{\columnwidth}
\begin{lstlisting}[style=contract_pseudo, frame=single, escapeinside={(*}{*)},
caption=Pseudocode of the domain contract., label=domainContract]
contract DomainContract:
  uint public thresholdT;
  certificate public certData;
  mapping(address => bool) public authorizedCAs;
  mapping(address => uint) public compensations;
  mapping(address => bytes) public certPubNonces;
  uint public certPubNoncesCount = 0;
  bool public allCertNonces = false;
  event allCertNoncesGathered();
  ...
  
  function DomainContract(params):
    thresholdT = params._thresholdT
    certData = params._certData
    authorizedCAs = params._authorizedCAs
    compensations = params._compensations

  function sendCertPubNonce(certPubNonce):
    if sender(*$\in$*)authorizedCAs:
      certPubNonces.push(certPubNonce)
      certPubNoncesCount++
      if certPubNonces == thresholdT:
        new event(allCertNoncesGathered)
        allCertNonces = TRUE

  function sendCertSignature(certSignature):
    if sender(*$\in$*)authorizedCAs & allCertNonces:
      certSigs.push(certSignature)
      certSigsCount++
      pay(sender, compensations[sender])
      if certSigsCount == thresholdT:
        new event(allCertSignaturesGathered)
  ...
\end{lstlisting}
\end{minipage}
\end{figure}

\begin{figure}[h]
\centering
\begin{minipage}[b]{\columnwidth}
\begin{lstlisting}[style=contract_pseudo, frame=single, escapeinside={(*}{*)},
caption=Pseudocode of the certificate verification process.,
label=cert_verification]
struct certificate = {
  subjectName,
  issuers[],
  notBefore,
  notAfter,
  publicKey,
  signature,
  ...
}

function verify(domainCert):

  cert = new certificate(domainCert.transaction)
  block = findBlock(domainCert.blockNo)
  root = block.getTransactionRoot()
  incl = domainCert.inclusionProof

  combinedKey = 1
  
  for CA in cert.issuers:
    if CA in trustedCAs:
      combinedKey *= CA.pubKey
  
  if(website.domainName == cert.subjectName &&
     time.now >= cert.notBefore &&
     time.now <= cert.notAfter &&
     sigVer(cert, combinedKey) &&
     proofMerkle(root, incl)):
    output: "accept certificate"
  else:
    output: "reject certificate"
\end{lstlisting}
\end{minipage}
\end{figure}

\end{document}